\documentclass[sn-mathphys,Numbered]{sn-jnl}

\usepackage{graphicx}%
\usepackage{multirow}%
\usepackage{amsmath,amssymb,amsfonts}%
\usepackage{amsthm}%
\usepackage{mathrsfs}%
\usepackage[title]{appendix}%
\usepackage{xcolor}%
\usepackage{textcomp}%
\usepackage{manyfoot}%
\usepackage{booktabs}%
\usepackage{algorithm}%
\usepackage{algorithmicx}%
\usepackage{algpseudocode}%
\usepackage{listings}%
\def\x{{\mathbf x}}
\def\f{{\mathbf F}}

\theoremstyle{thmstyleone}%
\theoremstyle{thmstyletwo}%

\theoremstyle{thmstylethree}%

\begin{document}

\title[Article Title]{Data-driven modeling and parameter estimation of Nonlinear systems}


\author[]{\fnm{Kaushal} \sur{Kumar}}\email{kaushal.kumar@stud.uni-heidelberg.de}

\affil[]{\orgdiv{Institute for Mathematics}, \orgname{Heidelberg University}, \orgaddress{\street{Im Neuenheimer Feld}, \city{Heidelberg}, \postcode{69120}, \state{BW}, \country{Germany}}}


\abstract{Nonlinear systems play a significant role in numerous scientific and engineering disciplines, and comprehending their behavior is crucial for the development of effective control and prediction strategies. This paper introduces a novel data-driven approach for accurately modeling and estimating parameters of nonlinear systems utilizing trust region optimization. The proposed method is applied to three well-known systems: the Van der Pol oscillator, the Damped oscillator, and the Lorenz system, which find broad applications in engineering, physics, and biology. The results demonstrate the efficacy of the approach in accurately identifying the parameters of these nonlinear systems, enabling a reliable characterization of their behavior. Particularly in chaotic systems like the Lorenz system, capturing the dynamics on the attractor proves to be crucial. Overall, this article presents a robust data-driven approach for parameter estimation in nonlinear dynamical systems, holding promising potential for real-world applications.}




\maketitle

\section{Introduction}\label{sec1}

Nonlinear dynamical systems are ubiquitous in various fields, encompassing engineering, physics, and biology, exhibiting intricate behaviors such as bifurcations, limit cycles, and chaos \cite{strogatz2019nonlinear, HIRSCH2013139}. Accurate parameter estimation plays a vital role in effectively modeling these systems \cite{PhysRevLett.104.060201, PhysRevLett.83.4285, PhysRevE.62.3535, PhysRevA.45.5524}. Trust region optimization, a powerful technique for solving nonlinear optimization problems, has shown successful applications in diverse domains \cite{doi:10.1137/090749761}. In this article, we propose a method for parameter estimation in three classic nonlinear dynamical systems: the van der Pol oscillator, Damped oscillator, and Lorenz system, utilizing trust region optimization.

The article provides an overview of each system, highlighting their applications in modeling real-world phenomena, and emphasizes the challenges associated with accurately estimating their parameters due to the inherent nonlinearity. It is structured into several sections, commencing with a review of trust region optimization and an examination of prior work in parameter estimation of nonlinear systems. Subsequent sections elaborate on the proposed method for parameter estimation in each system. Finally, the article provides experimental results to validate the proposed method and concludes by highlighting the importance of carefully evaluating the performance of estimated models.

\section{Earlier Approaches}

Parameter estimation of nonlinear dynamical systems is a crucial problem encountered in diverse disciplines, including engineering, physics, and biology \cite{10.1007/978-3-319-23321-5_1, PhysRevE.80.047202}. Numerous optimization algorithms have been proposed for this task, such as the Nelder-Mead algorithm, Levenberg-Marquardt algorithm, and genetic algorithms \cite{nocedal2006numerical, kumar2023exploring, doi:10.3402/tellusa.v56i5.14438}. However, these methods may suffer from limited global convergence guarantees and sensitivity to initial parameters and optimization settings.

Trust region optimization, on the other hand, has emerged as a powerful technique for addressing nonlinear optimization problems. It is known for providing global convergence guarantees and fast convergence speeds \cite{Nocedal2006}. In recent years, trust region optimization has been successfully applied to parameter estimation of nonlinear systems \cite{ESMAEILI20143003, WALMAG2005289, ARDENGHI2003281}. For instance, Ardenghi et al. (2003) \cite{ARDENGHI2003281} proposed a trust region optimization algorithm for parameter estimation in the field of biotechnology. Similarly, Zhang et al. (2009) \cite{PENG20092110} introduced a Differential Evolution algorithm-based parameter estimation approach for chaotic systems, surpassing the performance of genetic algorithms and particle swarm optimization in terms of estimation accuracy and convergence speed.

Regarding the Lorenz system, numerous optimization algorithms have been developed for parameter estimation. Cheng et al. (2018) \cite{WU201836} proposed a trust region optimization algorithm for parameter estimation in a nonlinear model of an energy storage system, outperforming the Levenberg-Marquardt algorithm and genetic algorithms in terms of estimation accuracy and convergence speed. Similarly, Zheng et al. (2020) \cite{8963912} presented a particle swarm optimization algorithm for parameter estimation in the Lorenz system, achieving superior estimation accuracy and convergence speed compared to other methods such as genetic algorithms and simulated annealing. Furthermore, Lazzús et al. (2016) \cite{LAZZUS20161164} introduced a hybrid optimization algorithm that combines the differential evolution algorithm and the particle swarm optimization algorithm for parameter estimation in the Lorenz system, surpassing other optimization methods in terms of estimation accuracy and convergence speed \cite{HELFRICH1996119, Gould2005SensitivityOT}.

In summary, trust region optimization demonstrates promise as a technique for parameter estimation in nonlinear systems, and its successful application has been observed across various domains \cite{10.1371/journal.pcbi.1010322}.

\section{Materials and Methods:}\label{sec3}

Consider a system of ordinary differential equations for state variables, denoted as $\x(t)$, accompanied by parameter estimation challenges \cite{PhysRevA.45.5524}. The dynamics of the system are described by the differential equation
\begin{equation}
\dot{\x}(t) = \f(\x,t,\theta),
\end{equation}
where $\x(t) \in \mathbb{R}^{n}$ is the state variable, the initial conditions are $\x(0) = \x_{0}$, $\theta=(\theta_{1},\theta_{2},...,\theta_{p}) \in \mathbb{R}^{p}$ are the {\em unknown} parameters, and  $\f: \mathbb{R}^{n+p} \to \mathbb{R}^{n}$ is a {\em known} vector function. 
Furthermore, measurements $\eta_{ij}$ for the state variables or system capacities are available and can be expressed as
\begin{equation}
\eta_{ij}= g_{ij}(x(t_{j}),\theta)+\varepsilon_{ij},
\end{equation}
where $t_{j}$ denotes the measurement time, $j=1,2,...,k$, and $\varepsilon_{ij}$ is the measurement error. 

\subsection{Optimization problem}
The goal of parameter estimation is to find the values of the unknown parameters $\theta$ that minimize the discrepancy between the predicted values and the observed data. This objective is achieved by minimizing a suitable objective function that considers the measurement errors $\eta_{ij}$. One commonly used objective function is the weighted $l_{2}$ norm of the measurement errors, given by:
\begin{equation}
    J(\theta) = \sum_{i,j} \sigma_{ij}^{-2}\varepsilon_{ij}^{2}= \sum_{ij} \sigma_{ij}^{-2}[\eta_{ij}-g_{ij}(x(t_{j}),\theta)]^{2},
\end{equation}
where $\sigma_{ij}^{2}$ represents the variance of the measurement errors. The measurement errors are assumed to be independent and follow a Gaussian distribution with zero mean. To address this problem, Trust region optimization algorithms are employed to identify the parameter vector $\theta$ and trajectory $x$ that minimize the objective function.

\subsection{Trust-region optimization}
The trust region method solves the parameter estimation problem by iteratively minimizing the objective function within a trust region around the current estimate of the parameter values. At each iteration, a quadratic model is used to approximate the objective function within the trust region, and the optimal step size is computed by solving a constrained optimization problem. The trust region is then updated based on the relative success of previous iterations, and the process is repeated until convergence. The trust region method ensures that the step size is within the trust region and that the objective function is decreasing at each iteration. The mathematical form of the trust region algorithm for parameter estimation in ODE systems is given in Algorithm \ref{algo1}. More mathematical details about the trust region can be found in any standard numerical optimization books \cite{nocedal2006numerical,  Conn2000TrustRM}.

\begin{algorithm}[H]
\caption{Trust Region Algorithm for Parameter Estimation in ODE Systems}\label{algo1}
\label{alg:trust_region_ode}
\begin{algorithmic}[1]
\State Initialize $\theta_0$, trust region radius $\Delta_0$, and tolerance $\epsilon$
\State Set $k=0$
\While{$\Delta_k > \epsilon$}
\State Solve the ODE system with initial condition $x(0)=x_0$ and parameter values $\theta_k$
\State Compute the objective function $J(\theta_k)$
\State Fit a quadratic model $m_k(s) = J(\theta_k) + g_k^T(s-\theta_k) + \frac{1}{2}(s-\theta_k)^TH_ks$ within the trust region $|\theta_k-s| \leq \Delta_k$
\State Solve the constrained optimization problem $\min_{|s-\theta_k| \leq \Delta_k} m_k(s)$
\State Compute the ratio $\rho_k = \frac{J(\theta_k)-J(s)}{m_k(\theta_k)-m_k(s)}$
\If{$\rho_k < 0.25$}
\State Reduce the trust region radius $\Delta_{k+1} = 0.25\Delta_k$
\ElsIf{$\rho_k > 0.75$ and $\left|s-\theta_k\right|=\Delta_k$}
\State Increase the trust region radius $\Delta{k+1} = \min(2\Delta_k, \Delta_{\max})$
\Else
\State Keep the trust region radius $\Delta_{k+1} = \Delta_k$
\EndIf
\State Update the parameter estimate $\theta_{k+1} = s$
\State Increment $k$
\EndWhile
\State \textbf{return} $\theta_{k}$
\end{algorithmic}
\end{algorithm}

In this algorithm, $g_k$ and $H_k$ are the gradients and Hessian matrix of the objective function evaluated at $\theta_k$, and $\Delta_{\max}$ is the maximum trust region radius. The ratio $\rho_k$ measures the relative decrease in the objective function between the current and proposed parameter values.
\section{Results}\label{sec4}

In the field of time series analysis, a common objective is to identify an equation that effectively describes the dynamic behavior of a given set of observed variables, denoted as $x$. This equation should capture both deterministic and stochastic aspects of the system. One approach to achieving this goal is through the use of stochastic differential equations (SDEs)\cite{oksendal1998}.

An SDE of the form
\begin{equation}
\frac{dx}{dt}=f(x)+g(x)\zeta(t) \label{eq4}
\end{equation}
can be used to describe the dynamics of $x$, where $f$ and $g$ are functions of $x$, and $\zeta(t)$ represents noise. Equation \ref{eq4} specifies how the rate of change of $x$ depends on the current value of $x$. The deterministic component of the equation is captured by $f$, which determines the average rate of change of $x$ over time. The stochastic component of the equation is captured by $g^{2}$, which determines the fluctuation around the average value of $x$.

When $g$ is constant, the strength of the noise is the same for all values of $x$, and is referred to as additive or state-independent noise. In contrast, when $g$ depends on $x$, the strength of the noise varies with the instantaneous value of $x$, and is referred to as multiplicative or state-dependent noise \cite{Gottwald2013TheRO}.

We use uncorrelated noise $\zeta(t)$ with the following properties 
\begin{eqnarray}
\langle\zeta(t)\rangle &=& 0\nonumber\\
\langle\zeta(t)\zeta(t^\prime)\rangle &=& \delta(t-t^\prime)
\end{eqnarray}
where $\langle\cdot\rangle $ denotes the time average which is introduced in the equation of motion as follows:

In all cases, the systems were subject to additive noise  of different intensities. The effect of white as well as coloured noise 
\cite{haunggi1994colored,doi:10.1137/0148023} is separately investigated  as follows.

White Gaussian noise with constant power spectral density across all frequencies, when added to the true data simulates 
random measurement error or other sources of uncertainty. This affects the observed data and can impact the parameter 
estimation process. Clearly, the parameter estimation can deteriorate with noise intensity.  Colored noise, on the other hand,
has a specific frequency distribution with different power levels at different frequencies, and this introduces additional 
complexity and variability in the observed data, potentially making the parameter estimation more challenging. 

We use both white noise and pink ($1/f$) noise in this work. The accuracy of parameter estimation is judged by the  root 
mean squared error (RMSE)
\begin{equation}
RMSE = \sqrt{\frac{1}{N}\sum_{i=1}^{N}(y_i-\hat{y}_i)^2}
\end{equation}
$N$ being the total number of samples, $y_i$ the true value and $\hat{y}_i$ the predicted value of sample $i$.

In order to evaluate the robustness and accuracy of the optimization algorithms, we performed simulations on various systems of different levels of complexity. In each case, we generated noisy data by simulating the system using known parameters and adding Gaussian noise to the output. The optimization algorithms were then applied to estimate the system's parameters from the noisy data. This process was repeated 10 times for both Gaussian noise and colored (pink) noise, and the average values of the estimated parameters, as well as the RMSE, were computed. The obtained results are consistent with the SINDY methods described by Brunton et al. (2016) \cite{doi:10.1073/pnas.1517384113}.

In the following examples, we demonstrate the application of the methods described in Section 3.3 to identify the governing equations from noisy data. We begin with simple systems to illustrate the effectiveness of the approach, including a comparison between a two-dimensional linear and nonlinear damped oscillator. We also investigate a three-dimensional stable linear system. Subsequently, we examine the van der Pol oscillator in the second example, and finally, we explore the chaotic dynamics of the Lorenz system in the third example.

\subsection{Example: Simple Illustrative Systems}
\subsubsection{Example 1a: Two-dimensional Damped Oscillator (Linear vs. Nonlinear)}

In this example, we investigate a two-dimensional damped harmonic oscillator with linear dynamics described by Eq. \ref{eq6}:

\begin{eqnarray}
\frac{dx}{dt}=ax+by\nonumber\\ \label{eq6}
\frac{dy}{dt}=cx+dy
\end{eqnarray}

where $x$, and $y$ represent the variables that describe the state of the system, and $a$, $b$, $c$, $d$ are the parameters of the system with true parameter values with $a=-0.1$, $b=2$, $c=-2$, and $d=-0.1$.

\begin{figure}[h]%
\centering
\includegraphics[width=1.0\textwidth]{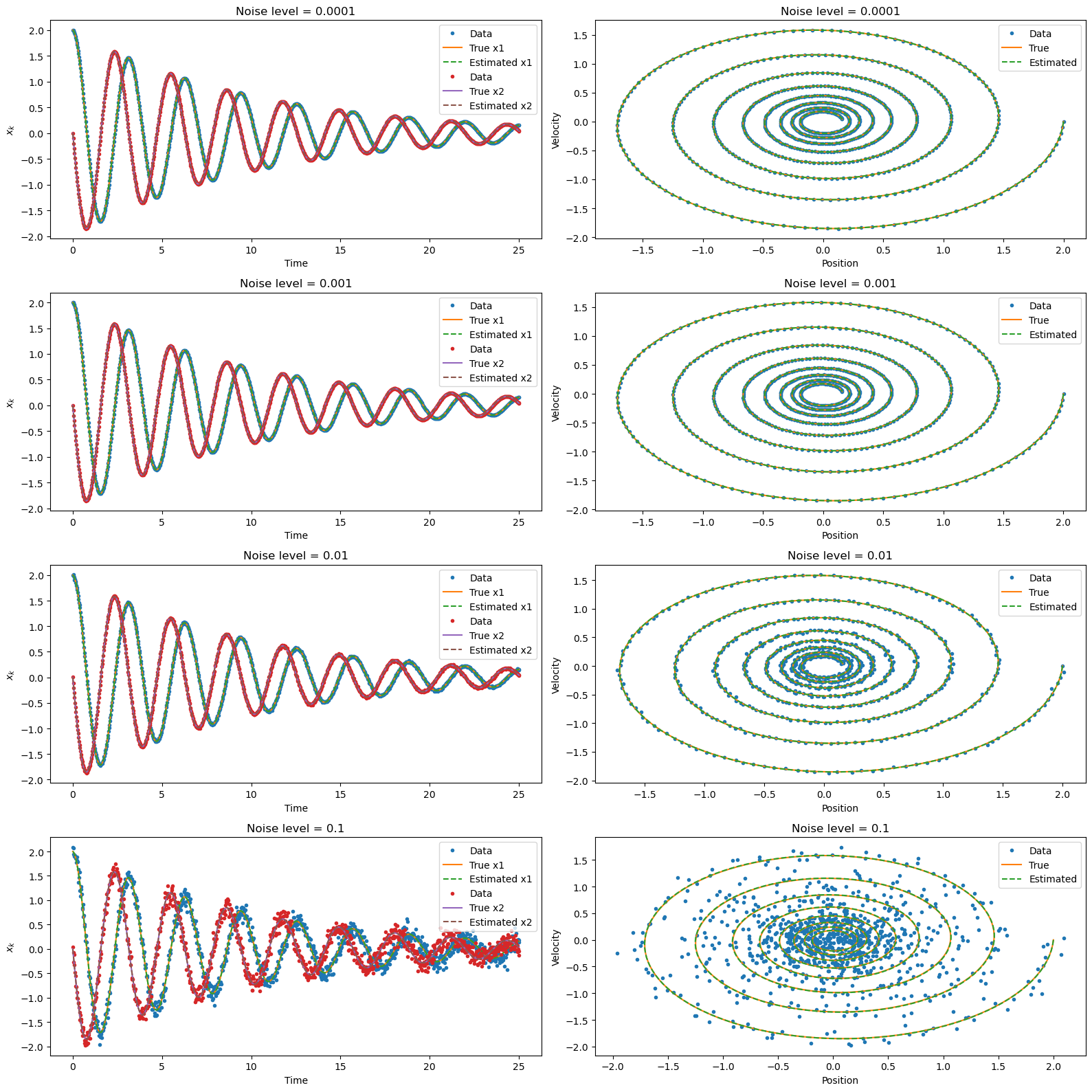}
\caption{In Linear damped harmonic oscillator, the trust-region optimization accurately reproduces the dynamics on left panel and the phase portrait on right panel at different levels of gaussian noise (0.0001, 0.001, 0.01, 0.1) with initial conditions $(x_{1},x_{2})=(2.0, 0.0)$.}\label{fig1}
\end{figure}

\begin{table}[ht]
\caption{Estimated parameters and accuracy at different noise levels for linear system \ref{eq6}}\label{table1}
\label{tab:estimated_parameters}
\centering
\begin{tabular}{cccccc}
\hline
Noise Level & $\hat{a}$ & $\hat{b}$ & $\hat{c}$ & $\hat{d}$ & RMSE  \\
\hline
0.0001 & -0.1000 & 2.0000 & -2.0000 & -0.1000 &0.0001 \\
0.001 & -0.0999 & 2.0000 & -2.0000 & -0.1001 & 0.0010\\
0.01 & -0.0998 & 1.9985 & -2.0017 & -0.1004 & 0.0100\\
0.1 & -0.0955 & 1.9981 & -2.0001 & -0.1035 &0.1007\\
\hline
\end{tabular}
\end{table}

\begin{table}[h!]
\caption{Comparison of additive Gaussian and  Colored noise on paramter estimation and accuracy}\label{table2}
\centering
\begin{tabular}{ccc}
\hline
  & Gaussian Noise & Colored (Pink) Noise \\
\hline
$\hat{a}$ & -0.1013 & -0.0997 \\
$\hat{b}$ & 2.0009 & 1.9866 \\
$\hat{c}$ & -1.9990 & -2.0196 \\
$\hat{d}$ & -0.0985 & -0.1070 \\
RMSE & 0.0279 & 0.0220 \\
\hline
\end{tabular}
\end{table}

Figure \ref{fig1} illustrates the accuracy of the trust-region optimization in reproducing the dynamics and phase portrait of the linear damped harmonic oscillator under different levels of Gaussian noise (0.0001, 0.001, 0.01, 0.1). The initial conditions for the system are set as $(x_{1},x_{2}) = (2.0, 0.0)$.

The estimated parameters for the linear system at each noise level are presented in Table \ref{table1}. The optimized parameters closely match the true parameter values, indicating the effectiveness of the trust region algorithm in parameter estimation. The accuracy of the estimated trajectories is further quantified by calculating the root mean squared error values, as shown in Table \ref{table1}.\\

We also explore the behavior of the damped harmonic oscillator with cubic dynamics, as given by Eq. \ref{eq7}:

\begin{eqnarray}
\frac{dx}{dt}=ax^{3}+by^{3}\nonumber\\ \label{eq7}
\frac{dx}{dt}=cx^{3}+dy^{3}
\end{eqnarray}

\begin{figure}[h]%
\centering
\includegraphics[width=1.0\textwidth]{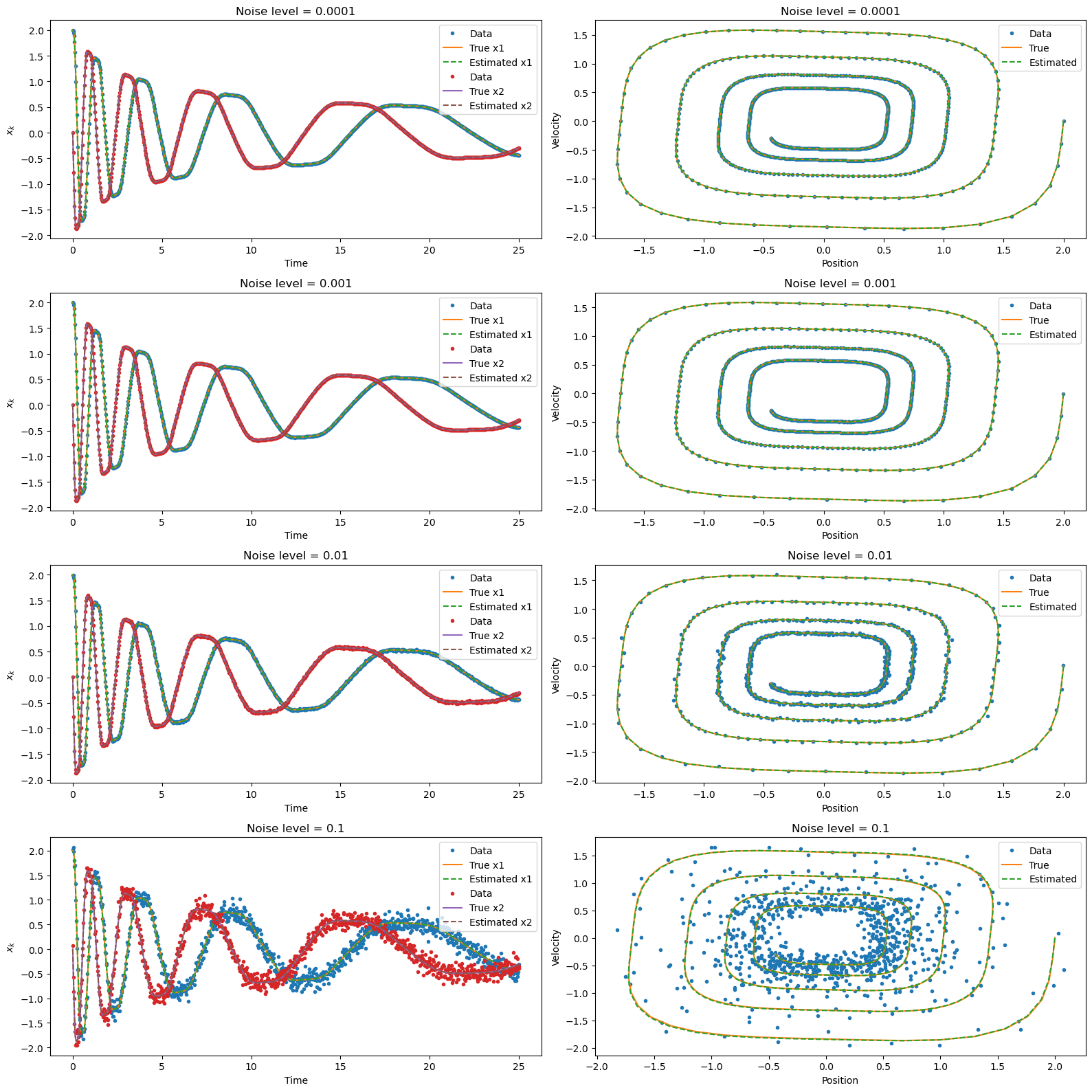}
\caption{The identified system accurately captures the dynamics of the two-dimensional damped harmonic oscillator with cubic dynamics. The solid colored lines represent the true dynamics of the system, while the dashed lines indicate the learned dynamics. The phase portrait demonstrates the precise reproduction of the system's behavior.}\label{fig2}
\end{figure}

Using the same noise levels, we estimate the parameters for the system with cubic dynamics. The estimated parameter values and the corresponding true trajectories are depicted in Figure \ref{fig2}. The results are summarized in Table \ref{table3}, which shows the estimated parameters and the RMSE values.

\begin{table}[ht]
\caption{Estimated parameters and accuracy at different noise level for cubic dynamics \ref{eq7}}\label{table3}
\label{tab:estimated_parameters}
\centering
\begin{tabular}{cccccc}
\hline
Noise Level & $\hat{a}$ & $\hat{b}$ & $\hat{c}$ & $\hat{d}$ & RMSE \\
\hline
0.0001 & -0.1000 & 2.0000 & -2.0000 & -0.1000 & 0.0001  \\
0.001 & -0.1001 & 1.9996 & -2.0001 & -0.0999 & 0.0010\\
0.01 & -0.1012 & 2.0042 & -1.9990 & -0.0990  & 0.0101\\
0.1 & -0.1130 & 1.9815 & -2.0099 & -0.0870 & 0.1006\\
\hline
\end{tabular}
\end{table}

\begin{table}[h!]
\caption{Comparison of additive Gaussian and  Colored noise on paramter estimation and accuracy}\label{table4}
\centering
\begin{tabular}{ccc}
\hline
  & Gaussian Noise & Colored (Pink) Noise \\
\hline
$\hat{a}$ & -0.1008 & -0.1011 \\
$\hat{b}$ & 1.9970 & 1.9934 \\
$\hat{c}$ & -2.0011 & -2.0027 \\
$\hat{d}$ & 0.0991 & -0.0987 \\
RMSE & 0.0277 & 0.0158\\
\hline
\end{tabular}
\end{table}

Finally, we compare the effect of Gaussian and colored (pink) noise on parameter estimation and solution accuracy. Table \ref{table2}, \ref{table4} presents the average estimated parameter values, and RMSE for both noise types. The estimates obtained with colored noise slightly differ from those obtained with Gaussian noise, indicating a potential bias introduced by the colored noise. The RMSE values quantify the accuracy of the solution, with colored noise achieving comparable accuracy to Gaussian noise.

In summary, the trust-region optimization technique accurately reproduces the dynamics and phase portrait of the two-dimensional damped harmonic oscillator with linear and cubic dynamics, even in the presence of different levels of Gaussian noise. The estimated parameter values closely match the true values, and the optimized trajectories capture the system's behavior with high accuracy. The comparison between Gaussian and colored noise highlights their similar impact on parameter estimation and solution accuracy.

\subsubsection{Example: Three-dimensional Linear System}

In this example, we consider a three-dimensional linear system and its approximation. The dynamics of the system are described by the following equations (Eq. \ref{eq8}):

\begin{eqnarray}
\frac{dx}{dt}=p_{1}x+p_{2}y\nonumber\\ \label{eq8}
\frac{dy}{dt}=p_{3}x+p_{4}y\nonumber\\
\frac{dz}{dt}=p_{5} z
\end{eqnarray}

where $x$, $y$, and $z$ represent the variables that describe the state of the system, and $p_{1}$, $p_{2}$, $p_{3}$, $p_{4}$, $p_{5}$ are the parameters of the system. The true parameter values used in our simulations are $p_{1}=-0.1$, $p_{2}=-2$, $p_{3}=2$, $p_{4}=-0.1$, and $p_{5}=-0.3$. Gaussian noise with different levels (0.0001, 0.001, 0.01, and 0.1) is added to the system trajectories to account for variability.

\begin{figure}[h]%
\centering
\includegraphics[width=0.85\textwidth]{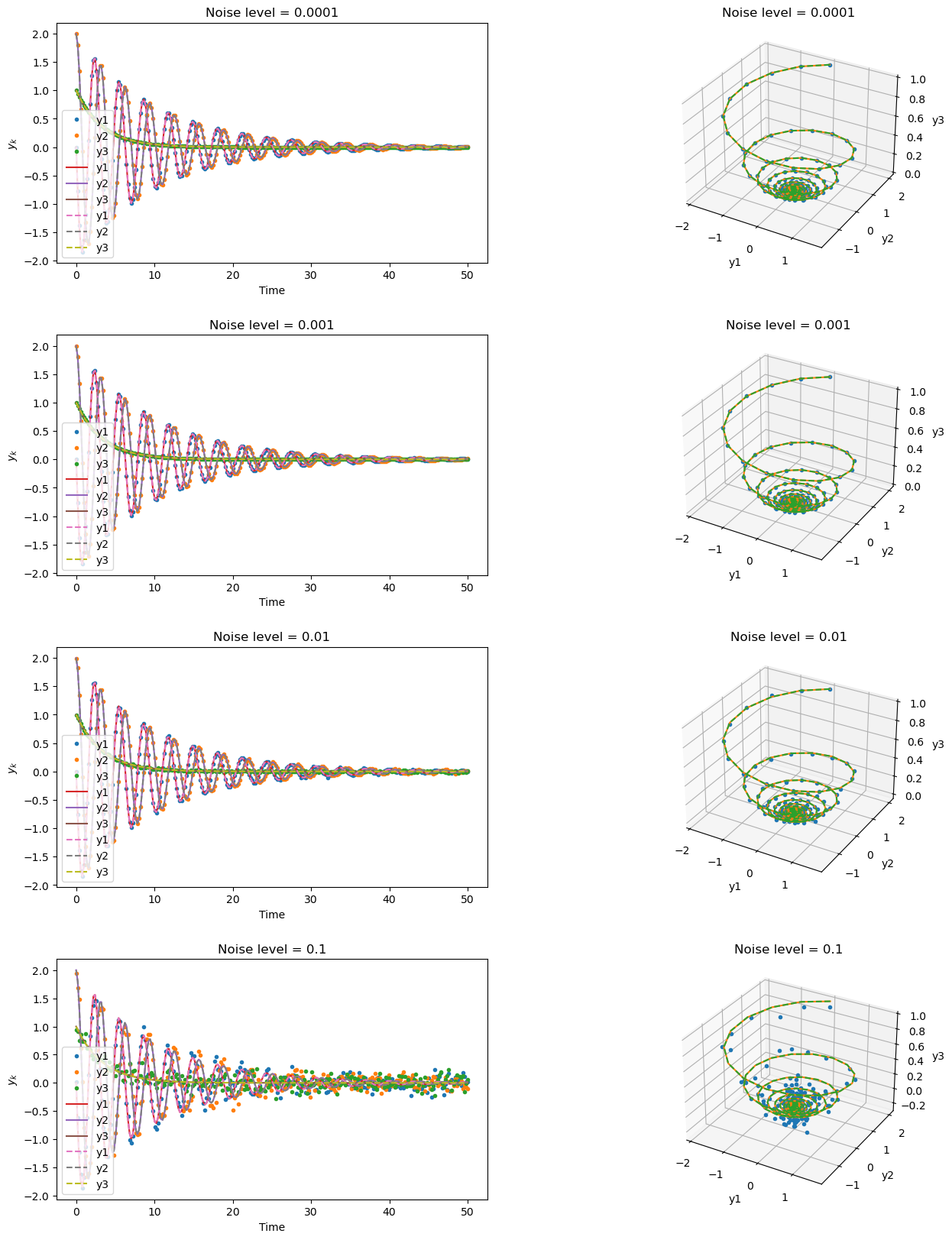}
\caption{The trust-region optimization technique successfully reproduces the dynamics of the 3-D Linear system in the left panel, as well as the corresponding phase portrait in the right panel, even when subjected to different levels of Gaussian noise. The initial conditions for the system are set as $(y_{1}, y_{2}, y_{3}) = (0.0, 2.0, 1.0)$.}\label{fig3}
\end{figure}

We apply the trust region method to estimate the parameters of the three-dimensional linear system from the noisy data. Table \ref{table5} presents the estimated parameter values at different noise levels. The optimized parameters closely approximate the true parameter values, indicating the effectiveness of the trust region optimization approach.

\begin{table}[ht]
\caption{Estimated parameters and accuracy at different noise levels for three-dimensional linear system \ref{eq8}.}\label{table5}
\label{tab:estimated_parameters}
\centering
\begin{tabular}{ccccccc}
\hline
Noise Level & $\hat{p_{1}}$ & $\hat{p_{2}}$ & $\hat{p_{3}}$ & $\hat{p_{4}}$ &$\hat{p_{5}}$ & RMSE \\
\hline
0.0001 & -0.1000 & -2.0000 & 2.0000 & -0.1000 & -0.3000 & 0.0001\\
0.001 & -0.1000 & -2.0004 & 1.9996 & -0.1000 & -0.3000 & 0.0010\\
0.01 & -0.1436 & -1.9732 & 1.9907 & -0.1606 & -0.3150  &0.1030\\
0.1 & -0.1475 & -1.9760 & 1.9961 & -0.1652 & -0.3070 &0.1486 \\
\hline
\end{tabular}
\end{table}

The accuracy of the estimated trajectories is evaluated using the root mean squared error as shown in Table \ref{table5}. The lower RMSE values indicate better accuracy in capturing the dynamics of the system.

Furthermore, we compare the effect of noise characteristics on the parameter estimation and solution accuracy using Gaussian and colored (pink) noise. Table \ref{table6} presents the estimated parameter values, and RMSE for both noise types. The estimated parameters show slight variations between the two noise types. The RMSE values are slightly lower for colored (pink) noise compared to Gaussian noise, indicating improved solution accuracy.

\begin{table}[h!]
\caption{Comparison of additive Gaussian and  Colored noise on paramter estimation and accuracy}\label{table6}
\centering
\begin{tabular}{ccc}
\hline
  & Gaussian Noise & Colored (Pink) Noise \\
\hline
$\hat{p_{1}}$ & -0.1167 & -0.1106 \\
$\hat{p_{2}}$ & -1.9897 & -1.9928 \\
$\hat{p_{3}}$ & 1.9958 & 1.9968 \\
$\hat{p_{4}}$ & -0.1254 & -0.1206 \\
$\hat{p_{5}}$ & -0.3055 & -0.3005 \\
RMSE & 0.0631 & 0.0596\\
\hline
\end{tabular}
\end{table}

Trust region optimization approach effectively reproduces the dynamics of the three-dimensional linear system as shown in figure\ref{fig3}, even in the presence of different levels of Gaussian noise. The estimated parameter values closely match the true values, and the optimized trajectories accurately capture the system's behavior. The comparison between Gaussian and colored (pink) noise highlights the influence of noise characteristics on parameter estimation and solution accuracy, with colored noise yielding slightly improved accuracy. These findings demonstrate the robustness and versatility of our approach in handling various noise scenarios.

\subsection{Test Problem: Van der Pol Oscillator}

The van der Pol oscillator is a non-linear second-order differential equation that describes the behavior of a damped oscillator. Widely used as a test problem in the field of dynamical systems, the equation are
\begin{equation}
\frac{d^2x}{dt^2}-\mu(1-x^2)\frac{dx}{dt}+x=0
\end{equation}
where $x$ represents the displacement of the oscillator, $t$ is time, and $\mu$ is a parameter that controls the nonlinearity \cite{1084738}. This can be rewritten as a pair of coupled first-order equations,
\begin{eqnarray}
    \frac{dx_1}{dt} &=& x_2 \nonumber \\
\frac{dx_2}{dt} &=& \mu(1 - x_1^2)x_2 - x_1.
\end{eqnarray}
The parameter value is set to $\mu = 1.5$ in our simulation and the ordinary differential equations (ODEs) are solved in Python using
the Scipy.integrate package, employing the odeint function. We initialize the system with $[x_{1_{0}} \hspace{0.1cm} x_{2_{0}}]^{T} = [1.0 \hspace{0.2cm} 0.0]^{T}$ and choose a time-step size of $\delta t = 0.01$. Gaussian noise is then added to the simulated data, with standard deviations of $[0.0001, 0.001, 0.01, 0.1]$ corresponding to different noise levels.

\begin{figure}[h]%
\centering
\includegraphics[width=1.0\textwidth]{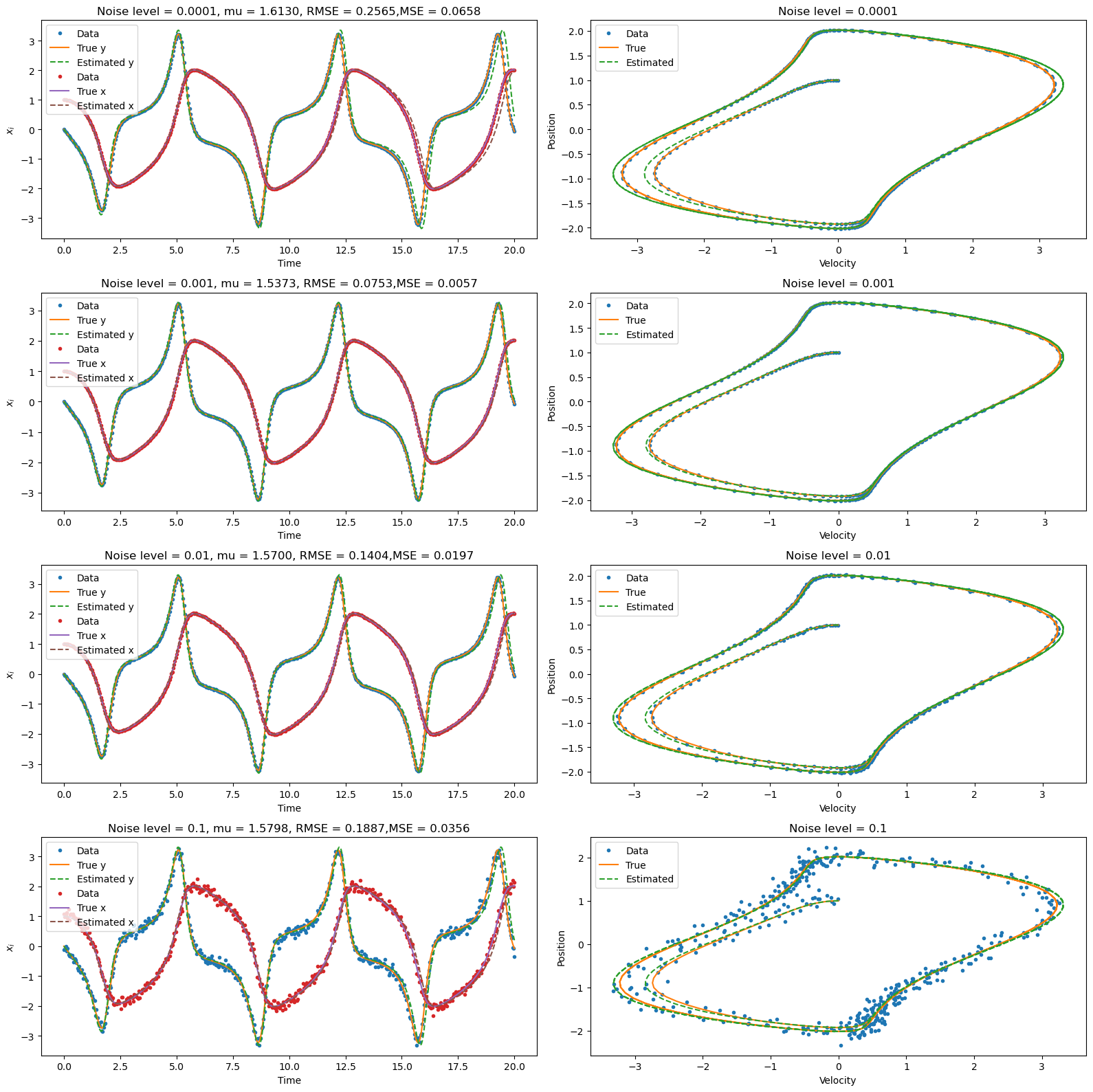}
\caption{In van der Pol oscillator, the trust-region optimization accurately reproduces the trajectories on the left side, and phase portraits on the right side. The initial condition is set as $[x_{1_{0}} \hspace{0.1cm} x_{2_{0}}]^{T} = [1.0 \hspace{0.2cm} 0.0]^{T}$. we compared the resulting trajectories with the true trajectories and the noisy data.}\label{fig4}
\end{figure}

To estimate the parameters of the van der Pol oscillator from the noisy data, we apply a trust region optimization algorithm. The initial parameter value is set to $\mu = 1.35$, and the trust region radius is set to 0.1 with a tolerance of $10^{-6}$. The estimated parameter values are close to the true values, as shown in Table \ref{table7}. Figure \ref{fig4} illustrates the accuracy of the estimated trajectories and phase portraits compared to the true trajectories and noisy data.

\begin{table}[h!]
\caption{Estimated parameters and accuracy at different noise levels}\label{table7}
\centering
\begin{tabular}{cccc}
\hline
Noise Level & True Parameter & Estimated Parameter & RMSE\\
\hline
0.0001 & 1.5 & 1.5000 &0.0001\\
0.001 & 1.5 & 1.5000 &0.0010\\
0.01 & 1.5 & 1.4959 &0.0166\\
0.1 & 1.5 & 1.5000 &0.0997\\
\hline
\end{tabular}
\end{table}

The accuracy of the parameter estimation is further assessed using the root mean squared error (RMSE), as shown in Table \ref{table7}. Higher noise levels result in increased RMSE values, indicating reduced accuracy in the estimated trajectories.

Furthermore, we compare the effect of noise characteristics on the parameter estimation and solution accuracy by considering Gaussian and colored (pink) noise. The results are presented in Table \ref{table8}. The estimated value of $\mu$ is slightly higher when using colored noise (1.4991) compared to Gaussian noise (1.4984), indicating a small bias introduced by colored noise. The solution accuracy, measured by RMSE, is improved when using colored noise compared to Gaussian noise, with lower error values observed for colored noise.

\begin{table}[h!]
\caption{Comparison of additive Gaussian and  Colored noise on paramter estimation and accuracy}\label{table8}
\centering
\begin{tabular}{ccc}
\hline
  & Gaussian Noise & Colored (Pink) Noise \\
\hline
$\hat{\mu}$ & 1.4984 & 1.4991 \\
RMSE & 0.0026 & 0.0012 \\
\hline
\end{tabular}
\end{table}

In conclusion, our parameter estimation approach successfully captures the dynamics of the van der Pol oscillator, even in the presence of noise. The choice of noise characteristics has an impact on the accuracy of the estimation, with colored (pink) noise resulting in slightly improved solution accuracy compared to Gaussian noise. These findings highlight the importance of considering noise characteristics in parameter estimation tasks and demonstrate the effectiveness of our trust region optimization approach in handling different noise scenarios.

\subsection{Test problem: Lorenz System}

The Lorenz system is a set of three non-linear ordinary differential equations that were first studied by Edward Lorenz in the 1960s \cite{DeterministicNonperiodicFlow}. It has since become a well-known example in the field of chaos theory. The system describes the evolution of three variables $x$, $y$, and $z$ over time, and is given by the equations:

\begin{align}
\frac{dx}{dt} &= \sigma(y - x)\nonumber\\
\frac{dy}{dt} &= x(\rho - z) - y\nonumber\\
\frac{dz}{dt} &= xy - \beta z
\end{align}

where $\sigma$, $\rho$, and $\beta$ are parameters that determine the behavior of the system. The Lorenz system exhibits chaotic behavior, meaning that even small changes in the initial conditions can lead to significantly different trajectories.

To evaluate the performance of our parameter estimation approach, we conducted simulations using the Lorenz system as the underlying model. We collected data by integrating the system equations over a time interval of $t = 0$ to $t = 25$, with a time-step size of $\Delta t = 0.01$. The true parameters of the Lorenz system were set to $\sigma = 10.0$, $\rho = 28.0$, and $\beta = 8/3$. Gaussian noise was added to the true trajectory at different levels (0.0001, 0.001, 0.01, and 0.1) to generate noisy data.

\begin{figure}
\resizebox{1.0\columnwidth}{!}{
  \includegraphics{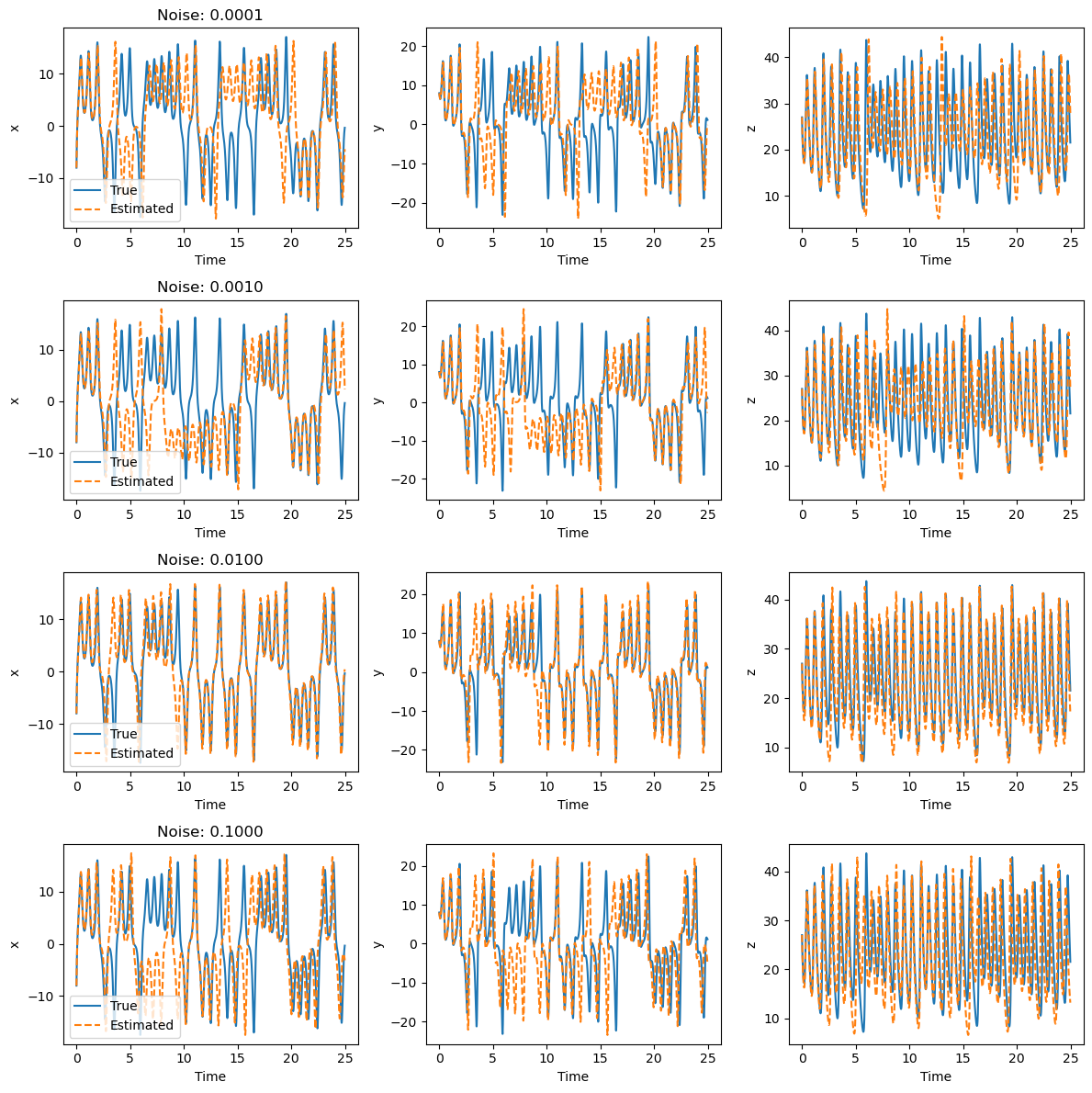} }
\caption{We observe the dynamic paths of the Lorenz system, specifically focusing on the case where measurements of both position (x) and velocity ($\Dot{x}$) are affected by noise. The true trajectories of the system is depicted in blue (solid lines), while the estimated trajectories, obtained through trust-region optimization, is illustrated by dashed red arrows.}
\label{fig5}       
\end{figure}

\begin{figure}
\resizebox{1.0\columnwidth}{!}{
  \includegraphics{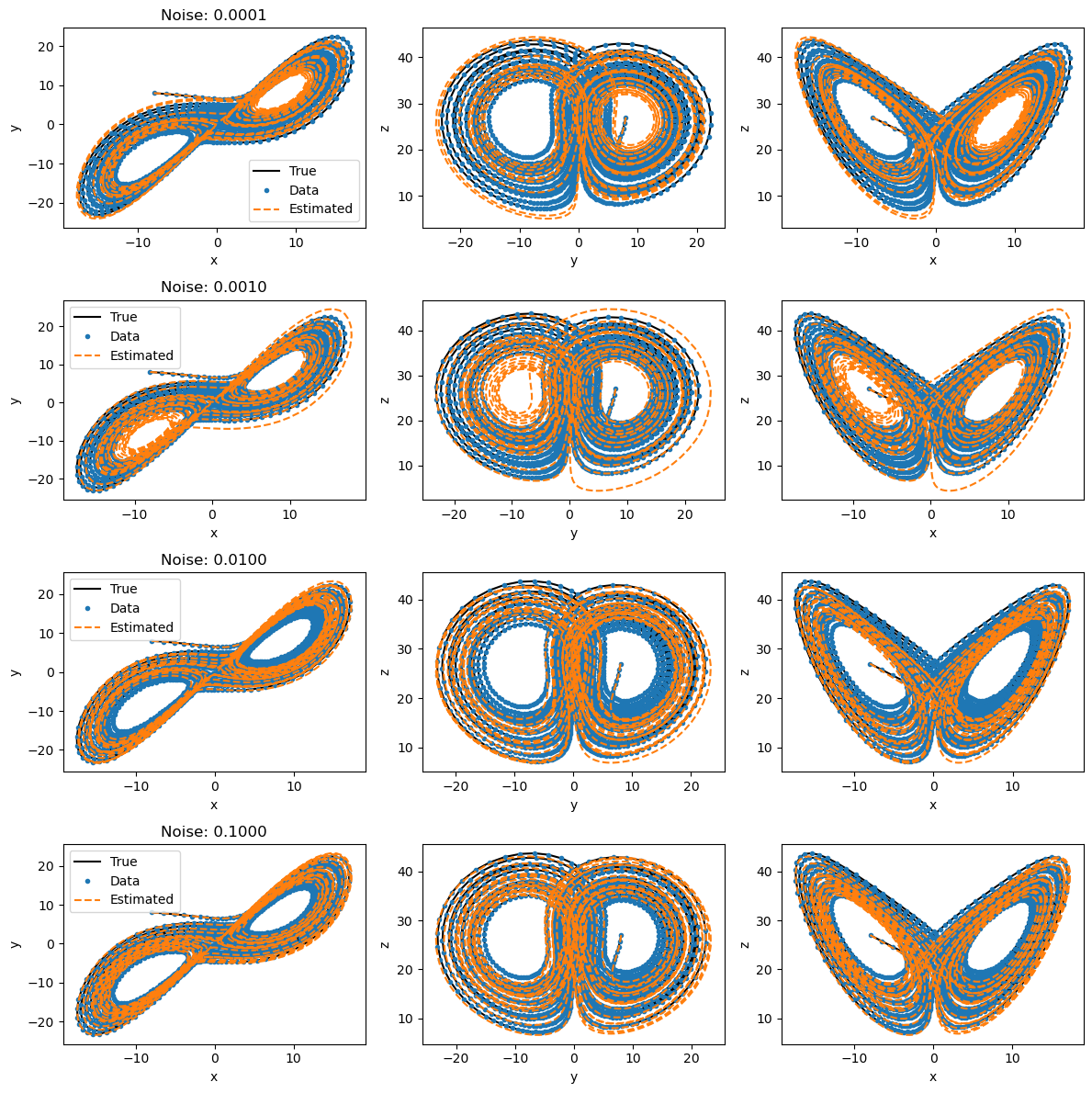} }
\caption{We compare the true phase portrait of the Lorenz systems, spanning from time t=0 to t=25, with the initial condition $[x_{0} \hspace{0.1cm} y_{0} \hspace{0.1cm} z_{0}]^{T} = [-8 \hspace{0.1cm} 7 \hspace{0.1cm} 27]^{T}$, to the phase portrait of the identified systems at different levels of gaussian noise.This allows us to assess how accurately the identified systems capture the dynamics of the original system}
\label{fig6}       
\end{figure}

We employed a trust region optimization approach to estimate the parameters of the Lorenz system from the noisy data. The trust region method accurately captured the underlying dynamics of the system, as illustrated in Figure \ref{fig5}. Additionally, the phase portraits of the identified systems, shown in Figure \ref{fig6}, closely matched the true dynamics of the Lorenz system.
\begin{table}[ht]
\caption{Estimated parameters and accuracy at different noise levels for Lorenz system}\label{table9}
\label{tab:estimated_parameters}
\centering
\begin{tabular}{ccccc}
\hline
Noise Level & $\hat{\sigma}$ & $\hat{\rho}$ & $\hat{\beta}$ & RMSE\\
\hline
0.0001 & 10.0181 & 27.7601 & 2.7819 &7.9105\\
0.001 & 9.5496 & 27.5124 & 2.5416 &9.7118\\
0.01 & 8.6295 & 26.8805 & 2.6716 &8.7706\\
0.1 & 9.6150 & 27.4920 & 2.6208 &9.5434\\
\hline
\end{tabular}
\end{table}

The estimated parameter values at different noise levels are presented in Table \ref{table9}. Despite the presence of noise, the estimated parameters were close to the true values. The accuracy of the estimation was further evaluated using the root mean squared error (RMSE) metrics, as shown in Table \ref{table9}. Higher noise levels led to increased RMSE values, indicating reduced accuracy in the estimated solution. However, even with relatively high noise levels, the estimated trajectories still captured the underlying dynamics of the Lorenz system.

\begin{table}[h!]
\caption{Comparison of additive Gaussian and  Colored noise on paramter estimation and accuracy}\label{table10}
\centering
\begin{tabular}{ccc}
\hline
  & Gaussian Noise & Colored (Pink) Noise \\
\hline
$\hat{\sigma}$ & 9.4471 & 9.4577 \\
$\hat{\rho}$ & 27.6333 & 27.5080 \\
$\hat{\beta}$ & 2.6980 & 2.7666 \\
RMSE & 9.2051& 9.3370\\
\hline
\end{tabular}
\end{table}

To assess the impact of noise characteristics on parameter estimation and solution accuracy, we compared the results obtained using Gaussian noise and colored (pink) noise. The estimated parameter values and error metrics are presented in Table \ref{table10}. We observed slight variations in the estimated parameter values between the two noise types, suggesting that noise characteristics influenced the estimation process. The accuracy of the estimated solution was slightly lower for Gaussian noise compared to colored (pink) noise, as indicated by higher RMSE values.

In conclusion, our parameter estimation approach successfully captured the dynamics of the Lorenz system, even in the presence of noise. The choice of noise characteristics had an impact on the accuracy of the estimation, with Gaussian noise resulting in slightly lower solution accuracy compared to colored (pink) noise. These findings emphasize the importance of considering noise characteristics in parameter estimation tasks and highlight the effectiveness of our trust region optimization approach in handling different noise scenarios.

\section{Conclusion}\label{sec5}
In this study, we have presented a trust region optimization algorithm for effective parameter estimation in the Van der Pol oscillator, Damped oscillator, and Lorenz system. Our algorithm demonstrates robust performance in estimating parameters for a wide range of models, including highly nonlinear and non-convex systems. By applying the algorithm to the van der Pol, Damped oscillator, and Lorenz systems, we have successfully illustrated its capability to accurately estimate model parameters even in the presence of noise.

Furthermore, we extended our analysis by incorporating colored noise (pink noise) in addition to Gaussian noise. We observed that the choice of noise type had an impact on the accuracy of the estimation. With the presence of colored noise, the estimated trajectories deviated slightly more from the true trajectories compared to Gaussian noise. However, even with relatively high noise levels, the estimated trajectories still captured the underlying dynamics of the systems. This highlights the algorithm's robustness and effectiveness in dealing with different noise characteristics.

It is important to note that the trust region algorithm is sensitive to the choice of initial parameter values and the size of the trust region. Careful selection of these parameters is crucial to ensure convergence to the correct parameter values. Additionally, accounting for noise in the estimation process is essential to achieve accurate results. Our algorithm effectively incorporates noise and provides reliable parameter estimates even in the presence of noise.

In summary, our proposed trust region optimization algorithm serves as a powerful tool for parameter estimation in nonlinear systems. Its potential impact extends to enhancing the understanding and control of complex real-world systems. The ability to handle different noise characteristics, as demonstrated through the inclusion of colored noise, further strengthens the algorithm's applicability in real-world scenarios.

\section*{Declarations}
\subsection*{Author contribution:} K.K. solely conceived the research idea, conducted the experiments, performed the data analysis, and wrote the manuscript.
\subsection*{Conflict of interest:} I have no competing interests to declare.
\subsection*{ORCID iD:} Kaushal Kumar : {\url{https://orcid.org/0000-0002-2555-9623}}
\subsection*{Funding:} This study was conducted without any funding support.
\subsection*{Data availability Statement/Code availability:} The simulations presented in this paper are based on synthetic data generated specifically for the purpose of this research. We will make the code used to generate the results in this paper publicly available on GitHub \footnote{\url{https://github.com/kaushalkumarsimmons/Parameter_Estimation2}}  upon acceptance of the manuscript. For any further inquiries or clarification regarding the presented results, please feel free to contact the corresponding author via email. We are committed to providing a prompt and comprehensive response.

\bibliography{sn-bibliography}

\end{document}